\documentclass{aastex62}

\usepackage{color}
\usepackage{graphicx}
\usepackage{subfigure}
\received{January 25, 2020}
\revised{February 16, 2020}
\accepted{\today}
\submitjournal{AJ}
\def\degree{${}^{\circ}$}
\shorttitle{PSF-NET}
\shortauthors{Peng et al.}
\begin{document}

\title{PSF--NET: A Non-parametric Point Spread Function Model for Ground Based Optical Telescopes}

\author[0000-0001-6623-0931]{Peng Jia}
\affil{College of Physics and Optoelectronics,  Taiyuan University of Technology, Taiyuan, 030024, China}
\affiliation{Key Laboratory of Advanced Transducers and Intelligent Control Systems, Ministry of Education and Shanxi Province, Taiyuan University of Technology, Taiyuan, 030024, China}
\affiliation{Department of Physics, Durham University, South Road, Durham DH1 3LE, UK}
\email{robinmartin20@gmail.com}

\author{Xuebo Wu}

\author{Huang Yi}

\author{Bojun Cai}

\author{Dongmei Cai}
\affiliation{College of Physics and Optoelectronics,  Taiyuan University of Technology, Taiyuan, 030024, China}

%% Note that the \and command from previous versions of AASTeX is now
%% depreciated in this version as it is no longer necessary. AASTeX 
%% automatically takes care of all commas and "and"s between authors names.

%% AASTeX 6.2 has the new \collaboration and \nocollaboration commands to
%% provide the collaboration status of a group of authors. These commands 
%% can be used either before or after the list of corresponding authors. The
%% argument for \collaboration is the collaboration identifier. Authors are
%% encouraged to surround collaboration identifiers with ()s. The 
%% \nocollaboration command takes no argument and exists to indicate that
%% the nearby authors are not part of surrounding collaborations.

%% Mark off the abstract in the ``abstract'' environment. 
\begin{abstract}
Ground based optical telescopes are seriously affected by atmospheric turbulence induced aberrations. Understanding properties of these aberrations is important both for instruments design and image restoration methods development. Because the point spread function can reflect performance of the whole optic system, it is appropriate to use the point spread function to describe atmospheric turbulence induced aberrations. Assuming point spread functions induced by the atmospheric turbulence with the same profile belong to the same manifold space, we propose a non-parametric point spread function -- PSF-NET. The PSF-NET has a cycle convolutional neural network structure and is a statistical representation of the manifold space of PSFs induced by the atmospheric turbulence with the same profile. Testing the PSF-NET with simulated and real observation data, we find that a well trained PSF--NET can restore any short exposure images blurred by atmospheric turbulence with the same profile. Besides, we further use the impulse response of the PSF-NET, which can be viewed as the statistical mean PSF, to analyze interpretation properties of the PSF-NET. We find that variations of statistical mean PSFs are caused by variations of the atmospheric turbulence profile: as the difference of the atmospheric turbulence profile increases, the difference between statistical mean PSFs also increases. The PSF-NET proposed in this paper provides a new way to analyze atmospheric turbulence induced aberrations, which would be benefit to develop new observation methods for ground based optical telescopes.\\
\end{abstract}

%% Keywords should appear after the \end{abstract} command. 
%% See the online documentation for the full list of available subject
%% keywords and the rules for their use.

%% From the front matter, we move on to the body of the paper.
%% Sections are demarcated by \section and \subsection, respectively.
%% Observe the use of the LaTeX \label
%% command after the \subsection to give a symbolic KEY to the
%% subsection for cross-referencing in a \ref command.
%% You can use LaTeX's \ref and \label commands to keep track of
%% cross-references to sections, equations, tables, and figures.
%% That way, if you change the order of any elements, LaTeX will
%% automatically renumber them.
%%
%% We recommend that authors also use the natbib \citep
%% and \citet commands to identify citations.  The citations are
%% tied to the reference list via symbolic KEYs. The KEY corresponds
%% to the KEY in the \bibitem in the reference list below. 
\section{Introduction}
For ground based large aperture optical telescopes, aberrations brought by thermal or gravity deformations can be reduced by the active optics system \citep{Su2004}. Normally the atmospheric turbulence induced aberration is the main limitation of  the performance of ground based optical telescopes. The imaging process of the ground based optical telescope can be modeled by:
\begin{equation} \label{eq:equation1}
Img(x,y)=[Obj(x,y)\ast PSF(x,y)]_{pixel(x,y)}+Noise(x,y),
\end{equation}
where $Obj(x,y)$ and $Img(x,y)$ are original and observed images. $PSF(x,y)$ is the point spread function (PSF) of the whole optical system, the variation of which is mainly caused by atmospheric turbulence induced aberrations. $[]_{pixel(x,y)}$ stands for the pixel response function of the detector and $Noise(x,y)$ stands for the noise from the background and the detector. Adaptive optics systems \citep{Babcock1953} and image restoration algorithms \citep{bertero1998} are two effective ways to reduce atmospheric turbulence induced aberrations. Conventional image restoration algorithms estimate $Obj(x,y)$ from $Img(x,y)$ with some prior information of $PSF(x,y)$, and adaptive optics systems try to keep the $PSF(x,y)$ small and stable during scientific observations.\\

The effectiveness of image restoration algorithms and that of adaptive optics systems rely on our understanding of the degradation process. For adaptive optics systems, because the atmospheric turbulence is continuous random medium that satisfies the frozen-flow assumption, we can obtain diffraction limited images in a small field of view if the adaptive optics system works in a relative high frequency \citep{Cortes2013}. With our increased understanding of  the turbulence spatial distribution and temporal variation, we can further improve the performance of multi--object adaptive optic systems \citep{Ono2016} or that of ground layer adaptive optics systems \citep{Li2020} through optimization of control or reconstruction strategies.\\

For image restoration algorithms, because image degradation caused by the atmospheric turbulence is a stochastic process, it will be hard to directly use atmospheric turbulence induced aberrations to analyze the degradation process. Since PSFs can reflect atmospheric turbulence induced aberrations, we can use PSFs to understand the imaging process in ground based optical telescopes. When the observation condition is stable (such as observations fed with AO systems) and the field of view is small enough to keep the PSF spatial invariant, direct deconvolution with PSFs obtained from reference stars can give good results \citep{Adorf1993, Magain1998,Starck2002}. For night--time seeing--limited long--exposure astronomical observations, because observation targets are mainly diffuse and point--like sources, blind deconvolution algorithms which estimate PSFs with some regularization conditions can increase image quality \citep{Bertero2000,Desider2009, Prato2013}. When the PSF is spatial variable, blind deconvolution of smaller partitioned images is effective and can improve astrometry and photometry accuracy \citep{Sun2013, Ciliegi2014, LaCamera2015, Jia2017}.\\

The development history of adaptive optic systems and that of image restoration algorithms mentioned above indicate us that if we want to further increase the performance of image restoration algorithms or that of adaptive optics systems, we need to further analyze properties of atmospheric turbulence induced aberrations. An appropriate PSF model is an important tool which should be updated as we obtain new knowledge or methods about the atmospheric turbulence. As we discussed above, since PSFs are introduced by the atmospheric turbulence, there exists a particular relation between the atmospheric turbulence and the PSF. In this paper, we assume PSFs induced by the atmospheric turbulence with the same profile belong to the same manifold space. PSFs generated by the Monte Carlo model can sample the manifold space. We take advantage of the great representation ability of  deep neural networks (DNNs) \citep{lecun2015,Goodfellow2016} and propose to train a non-parametric PSF model (PSF-NET)  with simulated PSFs. The trained PSF-NET can learn map between atmospheric turbulence profiles and PSFs.\\

The content of this article is organized in the following way. In section \ref{sec:frac}, we review classic PSF modeling methods and propose the concept of the PSF-NET. In section \ref{sec:cnn}, we test the PSF-NET with simulated data and show that the PSF-NET is an interpreted DNN. In section \ref{sec:analys}, we use the PSF-NET to analyze PSF variations brought by atmospheric turbulence profile variations. In section \ref{sec:con}, we will give our conclusions and propose our future research plans.\\
%%%%%%%%%%%%%%%%%%%%%%%%%%%%%%%%%%%%%
%--------------------------------------------------------------------
\section{The DNN based non-parametric PSF model: PSF-NET}\label{sec:frac}
\subsection{Classic PSF modeling method}
Analytical functions are used as PSF models firstly \citep{Moffat1969, Kormendy1973}. The standard point-symmetric model has been used for decades \citep{Stetson1992}. The Moffat function is a widely used analytical PSF model whose amplitude can be expressed as follows:
\begin{equation} \label{eq:equation2}
M_{A}(x,y)=\frac{A}{(1+x^{2}/\alpha^{2}_{x}+y^{2}/\alpha^{2}_{y})^{\beta}},
\end{equation}
with  $\alpha_{x}$, $\alpha_{y}$ and $\beta$ are positive real numbers. Moreover the condition $\beta > 1$ is imposed to ensure a finite integral of the function. The Moffat model can represent PSFs with some precision, but it is limited to long-exposure symmetric PSFs. Two different approaches are further proposed to model PSFs: modeling PSFs with combination of different bases \citep{Murtagh1995,Bernstein2002,Massey2005} or extracting PSF model directly from real observation data \citep{King1971, Lupton2001, Jarvis2004, Jee2007,Sun2017,Wang2018}. Because long exposure images or images generated by stacking several short exposure images can average stochastic properties of atmospheric turbulence, PSF modeling methods mentioned above are effective and have became standard methods for sky survey telescopes \citep{Chang2012,Li2016,Lu2018}.\\

For short exposure images or images obtained by telescopes with AO systems, PSFs have spatial and temporal variations brought by the atmospheric turbulence or the residual error. To maximize the scientific output, PSF reconstruction algorithms are proposed and the concept of them are estimating residue wavefront error from AO telemetry data \citep{Veran1997,Gendron2006}. Several analytical PSF reconstruction algorithms are proposed for different telescopes including on-axis PSF reconstruction \citep{Jolissaint2012}, off-axis PSF reconstruction \citep{Witzel2014} for ordinary AO systems, PSF reconstruction for ground layer adaptive optic system \citep{Peter2010,Villecroze2012}, multi-object adaptive optic system \citep{Martin2016} and multi-conjugate adaptive optic system \citep{Gilles2018}. These analytical PSF reconstruction algorithms can provide PSFs for image deconvolution \citep{Fusco2000, 2014Jia}. However, reconstructed PSFs provided by these algorithms need additional wavefront measurements and are effective only within the field of view of AO systems. \\
%--------------------------------------------------------------------
%%%%%%%%%%%%%%%%%%%%%%%%%%%%%%%%%%%%%
\subsection{ The concept of PSF-NET Modelling Method}
Besides PSF modeling methods discussed above, are there some other ways to model PSFs that do not require strict observation conditions or additional instruments? To answer this question, we need to review properties of the atmospheric turbulence and find a new way to model the PSF. The atmospheric turbulence is continuous medium which is hard to model in an analytical way. Right now, the Monte Carlo model is widely used for performance evaluation of adaptive optics systems \citep{Carbillet2005, Wang2010, Rigaut2013, Basden2018}. In a Monte Carlo model, we divide the extended atmospheric turbulence into several different layers. Each layer has different wind speed, wind direction and $C_n^2$ \citep{Roggemann1995}. Wavefront aberrations brought by each layer are modeled by thin phase screens which satisfy Kolmogorov or Von Karman power spectrum \citep{andrews2005}. Light from celestial objects travels through these layers according to Fresnel propagation law and generates PSFs of every instantaneous moment in the focal plane. With fixed exposure time, we can get PSFs through stacking all PSFs of instantaneous moments in the same position of the field of view.\\

The Monte Carlo model indicates us that the atmospheric turbulence is a complex stochastic model instead of a random model whose properties can not predict at all. Given an atmospheric turbulence Monte Carlo model, its state (such as the number of thin layers, the wind speed, the wind direction and the $C_n^2$ of each layer) is the same, albeit the state is sampled by different random numbers in atmospheric turbulence phase screens. The finite state of the atmospheric turbulence Monte Carlo model would generate PSFs with finite state. These PSFs belong to the same manifold space. If we have enough number of these PSFs to sample the manifold space, we could build a PSF model to represent the manifold space. Furthermore, if the Monte Carlo model and measurements of the state of the atmospheric turbulence are accurate in some degrees, the PSF model obtained from PSFs generated by the Monte Carlo model is an optimal representation of PSFs obtained from real observations.\\

\begin{figure}
   \centering
   \includegraphics[width=0.6\textwidth]{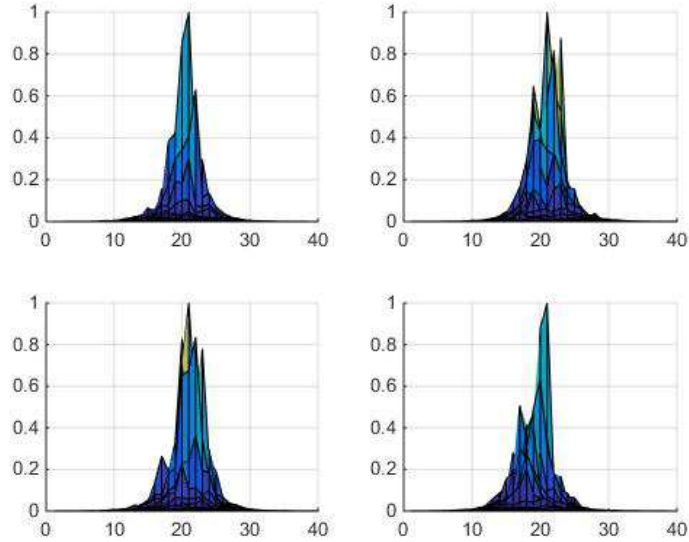}
      \caption{3D plot of four short exposure PSFs which have highly variable shapes.}
   \label{figure1}
   \end{figure}

Thanks to newly developed site testing instruments \citep{Shepherd2014,Osborn2018}, we can obtain high accuracy atmospheric turbulence profile measurements with high spatial and temporal resolution. At the same time, DNN has become complex enough to represent very complex function. For example, point spread function of images of extended objects obtained by telescopes with AO systems can be estimated with generative adversarial network \citep{long2020}. For short exposure images which have more complex PSFs, a generative model with DNN (Cycle-GAN) can restore any frames of short exposure solar images in a self-supervised way \citep{Jia2019}. During image restoration, the Cycle-GAN actually builds a model for short exposure PSFs, which have very complex structure as shown in figure \ref{figure1}. These works indicate us that contemporary DNNs are complex enough to model PSFs.\\

The development of site testing instruments, Monte Carlo simulation methods and DNNs have provided all necessary tools to build a non-parametric PSF model, which can reflect properties of PSF directly according to atmospheric turbulence profile measurements. With the concept mentioned above, we propose PSF-NET to learn map between the atmospheric turbulence profile and the PSF. We will discuss details of PSF-NET in the next subsection.\\
%--------------------------------------------------------------------
%--------------------------------------------------------------------

\subsection{The Structure and Training Strategy of the PSF-NET}
\subsubsection{The Structure of the PSF-NET}
Applications of the Cycle-GAN in short exposure image restoration \citep{Jia2019} indicate us the following key points in building an atmospheric turbulence induced PSF model:\\

1. CNN as complex as that in the Cycle-GAN is required to represent PSF models. Deeper CNN could model more complex PSFs, but it would cost much more computer resources.\\
2. The Cycle structure which learns the degradation and the restoration at the same time 
is efficient in training complex DNNs, because this structure could increase training speed and prevent over-fitting.\\
3. Extended images, such as images of galaxies or that of solar images, should be used to prevent over-fitting, because these images contain abundant effective information in different spatial scales.\\
4. Because PSFs generated by the Monte Carlo model are used to train the PSF model, these PSFs should be able to sample the manifold space of PSFs generated by the same turbulence profile. To fulfill this requirement, we need to carry out simulation with different random number to generate simulated PSFs.\\

\begin{figure}
   \centering
   \includegraphics[width=0.5\hsize]{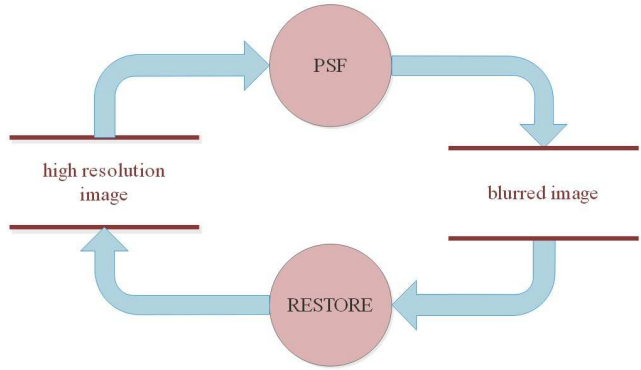}
      \caption{The overall structure of PSF-NET.The PSF network is used to model the PSF, which transforms high resolution images into blurred images. The RESTORE network is used to model deconvolution kernel, which transforms blurred images into high resolution images.}
   \label{figure2}
   \end{figure}

\begin{figure*}
   \centering
   \includegraphics[width=\hsize]{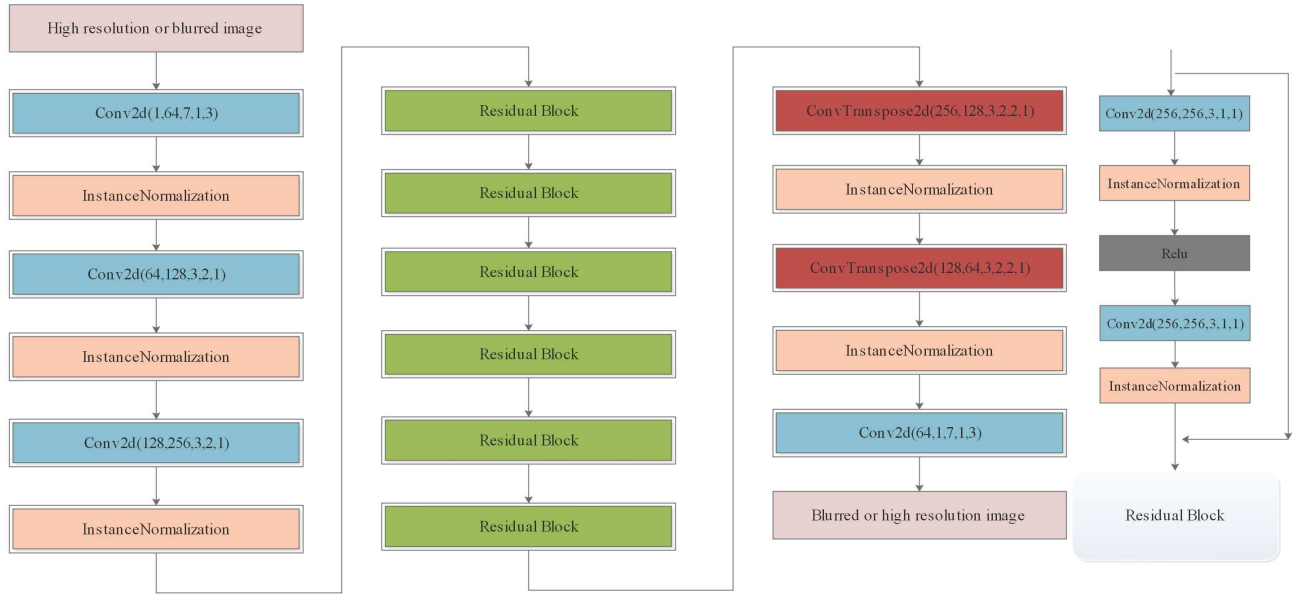}
      \caption{A structure diagram of a CNN, which could either be the PSF or the RESTORE. This CNN consists of convolutional layers (Conv2d in blue), instance normalization layers (IN in pink), Conv-transpose layers (ConvTranspose2d in red) and Residual blocks (Residual block in green).The structure of residual block is shown on the right. The Relu layer in gray color is used in the residual block as activation function. }
   \label{figure3}
   \end{figure*}

According to key points mentioned above, we propose to use two CNNs to form a Cycle-CNN structure as the PSF-NET. The structure of the PSF-NET is shown in figure \ref{figure2}. It includes two CNNs which have the same structure as that in the CycleGAN. Each CNN is a encoder-decoder DNN as shown in figure \ref{figure3}. An input image will pass through a few convolutional layers and instance normalization layers. Then 6 residual blocks are used to learn complex structure of PSFs. If we need to model more complex PSFs, we can simply add more residual blocks after these 6 residual blocks. The output of these residual blocks will be transmitted to several transpose convolution layers. After these layers, we can obtain output of the CNN.\\

 In the PSF-NET, one of the CNN (PSF neural network) learns the PSF. The other CNN (RESTORE neural network) learns the inverse function of the PSF (deconvolution kernel). The RESTORE neural network provides an effective way to evaluate the PSF estimated by the PSF-NET, because a trained RESTORE should be able to restore all images blurred by PSFs induced by the same turbulence profile. Because the PSF-NET is trained to learn the manifold of PSFs, pairs of images are required as the training set:  high resolution images and blurred images. High resolution images are images with complex structures, while blurred images are generated through convolution of original images and PSFs generated by the Monte Carlo model. We will discuss the training strategy of the PSF-NET in the following subsection.

%--------------------------------------------------------------------
%--------------------------------------------------------------------
%--------------------------------------------------------------------
\subsubsection{The Training strategy of the PSF-NET}
In this paper, PSFs are generated by Durham Adaptive Optics Simulation Platform \citep{Basden2018} with a high fidelity atmospheric turbulence phase screen generation method \citep{jia2015real,jia2015simulation}. Detailed parameters of our simulation are shown in table \ref{table1}. To assure that the manifold space is sampled well by PSFs generated by the Monte Carlo method, we carry out our simulation 20 times with different random numbers for each configuration.\\ 
%--------------------------------------------------------------------
	\begin{table}
	\caption{Parameters for Monte Carlo Simulation}                 
	\centering          
	\begin{tabular}{c c}     % 7 columns 
	\hline\hline       
    % To combine 4 columns into a single one 
	Name  &  Parameters\\
	\hline                    
	Diameter of Telescope & 1.0 metre\\ 
	Optical Design & Ideal Optical Telescope\\
	Field of View of each PSFs & 2.5 arcsec \\
	Field of View of our Simulation & 2.5 arcmin\\
	Exposure Time & 1 sec\\
	Pixel Scale & 0.01 arcsec\\
	Observation Band & 656.28 nm\\
	\hline                  
	\end{tabular}
	\label{table1}
	\end{table}
%--------------------------------------------------------------------

Besides, we convolve solar images observed by the New Vacuum Solar Telescope (NVST) in different wavelengths with simulated PSFs to train and test the PSF-NET \citep{liu2014}. All solar images are restored by the speckle reconstruction algorithm to the diffraction limit of the NVST \citep{li2015}. Solar images from H-alpha wavelength observed between 2018 April 2 and 3 which have filament structure are convolved with simulated PSFs as the training set, while solar images from TiO wavelength which have granulation structure observed on 2014 November 17 are convolved with PSFs generated with different random number as the test set. Because these images are observed in different wavelengths, they have different pixel scales (0.136 arcsec for TiO wavelength and 0.052 arcsec for H-alpha wavelength). However, all these images are convolved with PSFs generated in  656.28 $nm$ wavelength. We do not consider aberration or pixel scale variations caused by different wavelengths in this paper, because solar images from different wavelengths are only used to show generalization ability of the PSF-NET. The generalization of the PSF-NET should satisfy the following condition: the RESTORE neural network in the PSF-NET should be able to restore all images blurred by PSFs in the same manifold space, regardless of their contents and vice versa.\\

Since the PSF neural network is used to transform high resolution images to blurred images and the RESTORE neural network is used to transform blurred images to high resolution images, we need to design loss function according to this philosophy. First of all, the output of the PSF network should be the same as the original high resolution image and vice versa. It can be defined by the identity loss function as shown in equation \ref{eq:equation3},
%--------------------------------------------------------------------
\begin{equation}\label{eq:equation3}
L_{idt}=||PSF(Img_{High})-Img_{Blur}||_2 +\\
 ||RESTORE(Img_{Blur})-Img_{High}||_2,\\
\end{equation}
%--------------------------------------------------------------------
where $Img_{High}$ and $Img_{Blur}$ are high resolution images and blurred images, $|| ||_2$ stands for $L_2$ norm.\\

\begin{figure}
   \centering
   \includegraphics[width=0.4\textwidth]{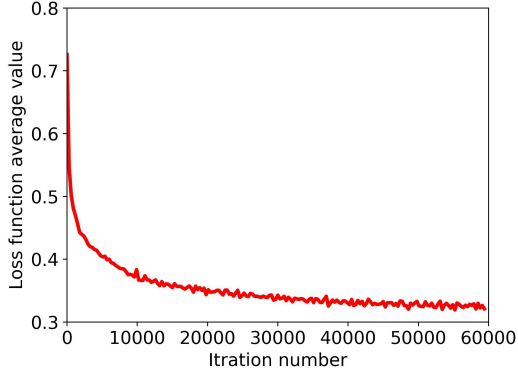}
      \caption{This figure shows loss function values after every 300 iterations. It is obvious that after 40000 iterations, values of loss function gradually converge.}
   \label{figure4}
   \end{figure}

Secondly, if we input the output of the PSF neural network into the RESTORE neural network, we should get the same or a similar image as the original blurred image, and vice versa, which could be defined by:\\
%-------------------------------------------------------------------- 
\begin{equation}\label{eq:equation4}
L_{rec}=||RESTORE(PSF(Img_{High}))-Img_{High}||_2+\\
||PSF(RESTORE(Img_{Blur}))-Img_{Blur}||_2.\\
\end{equation}
%--------------------------------------------------------------------
We add both of these two functions together as loss function to train the PSF-NET as defined below:\\
%--------------------------------------------------------------------
\begin{equation} \label{eq:equation5}
L = L_{idt}+L_{rec}.
\end{equation}
%--------------------------------------------------------------------
We use paired images to train the PSF-NET in a supervised way. The size of input images is $172 \times 172$ pixels. We randomly rotate pairs of these images by 0\degree, 90\degree, 180\degree or 270\degree and subtract minimal values for each image and normalize these images before we input them into the PSF-NET. We train the PSF-NET in a computer with 128 GB memory, two Nvidia GTX 1080 graphics cards and two Xeon E5 2650 processors. We set the batch size as 2 and use the Adam algorithm as optimization algorithm with learning rate of 0.0001 and beta as  (0.5, 0.999). It would cost about 17330s (60000 iterations) to train the PSF-NET in our computer. We plot average values of loss functions of every 300 iterations in figure \ref{figure4}. We can find that although values of loss functions fluctuate in each iteration, they will gradually converge after 40000 iterations. For each PSF-NET, we will train it for 60000 iterations.\\
%--------------------------------------------------------------------
%-------------------------------------------------------------------- 
%%%%%%%%%%%%%%%%%%%%%%%%%%%%%%%%%%%%%
\section{Applications of the PSF-NET}\label{sec:cnn}
\subsection{Data Set Preparing}
The PSF-NET learns the manifold of PSFs generated by atmospheric turbulence of the same profile. The turbulence profile includes four dimensions: height, $C_n^2$, wind speed and wind direction. We focus on variations of the height and the $C_n^2$ in this paper and set the wind speed and direction as the same fixed values for different turbulence profiles. For different turbulence profiles, we change the number of layers, the height and the $C_n^2$. Large amount of PSFs are generated by the Monte Carlo model with different turbulence profiles shown in table \ref{table2} and the same parameters defined in table \ref{table1}. We convolve these PSFs with high resolution solar images as the training set and the test set as shown in table \ref{table3}. \\
%--------------------------------------------------------------------
	\begin{table}
	\caption{Specific parameters of atmospheric turbulence}                 
	\centering          
	\begin{tabular}{c c}     % 2 columns 
	\hline\hline       
    % To combine 4 columns into a single one 
	Name  &  Parameters\\
	\hline                    
	Number of layers & 4\\ 
	Normalized integrated $C_n^2$ in different layers  & [ 0.30 , 0.20 , 0.30 , 0.20 ]\\
	Height in metre & [ 500 , 1000 , 2000 , 4000 ] \\
	\hline                  
	\end{tabular}
	\label{table2}
	\end{table}
%--------------------------------------------------------------------
%--------------------------------------------------------------------
	\begin{table}
	\caption{Data set used in this paper}                 
	\centering          
	\begin{tabular}{c c c c}     % 7 columns 
	\hline\hline                         
	Data Set &Wavelength&Number of images &Number of PSFs \\
	\hline   
	Training set & H-alpha & 2700 & 300  \\
	Test set & H-alpha & 900 & 50 \\  
	Test set & TiO & 900 & 50    \\  
	\hline   
	\end{tabular}
	\label{table3}
	\end{table}
%--------------------------------------------------------------------
\subsection{Test Generalization Ability of the PSF-NET}
After training the PSF-NET with the strategy discussed in subsection 2.3.2, we will first test generalization ability of the PSF-NET. The generalization ability of PSF-NET includes two main aspects: the PSF neural network can represent all PSFs generated by the atmospheric turbulence with the same profile, while the RESTORE neural network should be able to restore all images blurred by PSFs in the same manifold space. The first aspect is hard to test, but the second aspect is easy to test. Thanks to the Cycle-CNN structure of the PSF-NET, validation of the second aspect could provide evidence of the first aspect and we will test the second aspect in this part.\\

\begin{figure*}
\centering
\subfigure[High Resolution Image]{
\begin{minipage}[t]{0.2\linewidth}
\centering
\includegraphics[width=1\textwidth]{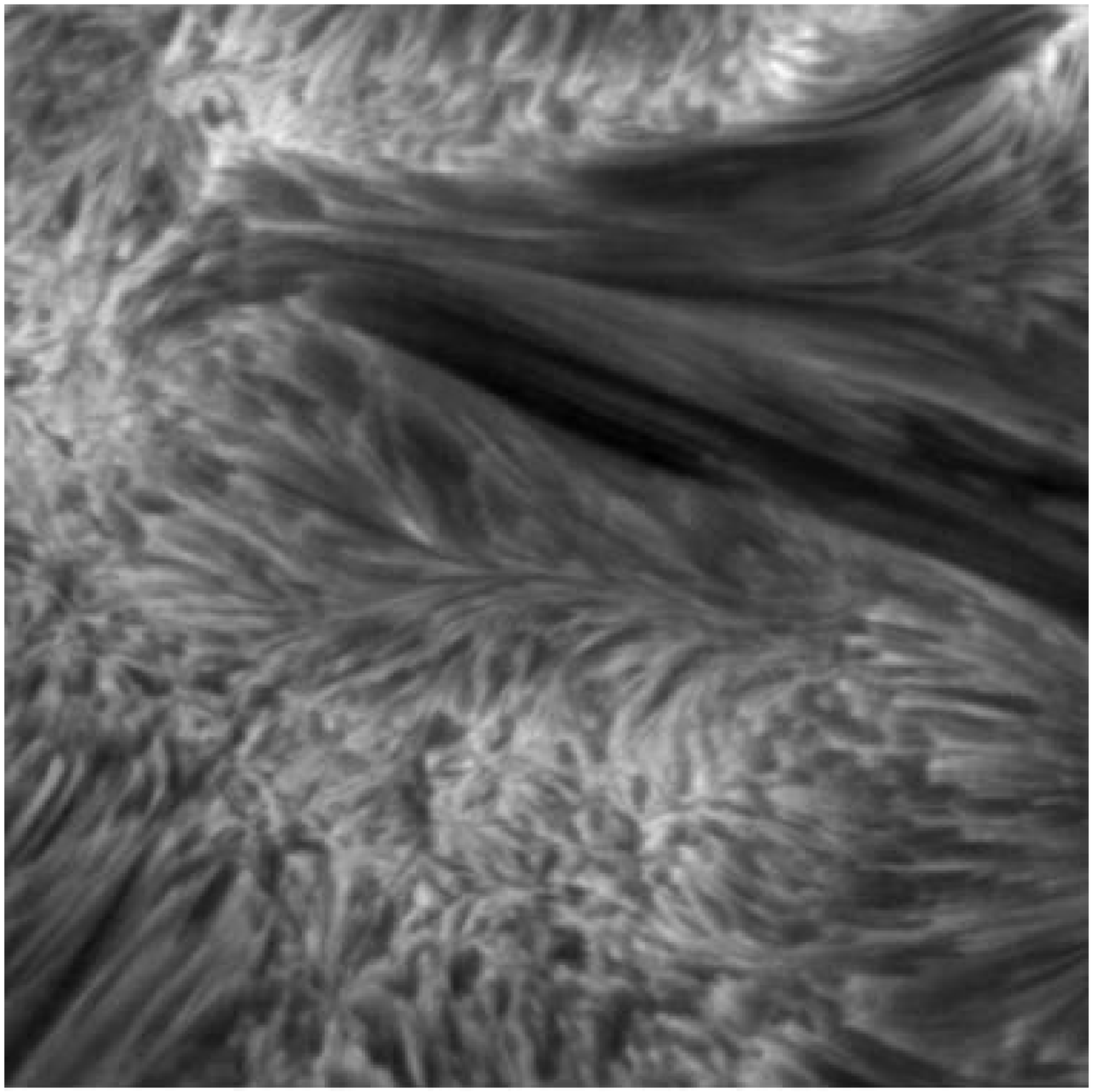}
%\caption{fig1}
\end{minipage}%
}%
\subfigure[Blurred Image]{
\begin{minipage}[t]{0.2\linewidth}
\centering
\includegraphics[width=1\textwidth]{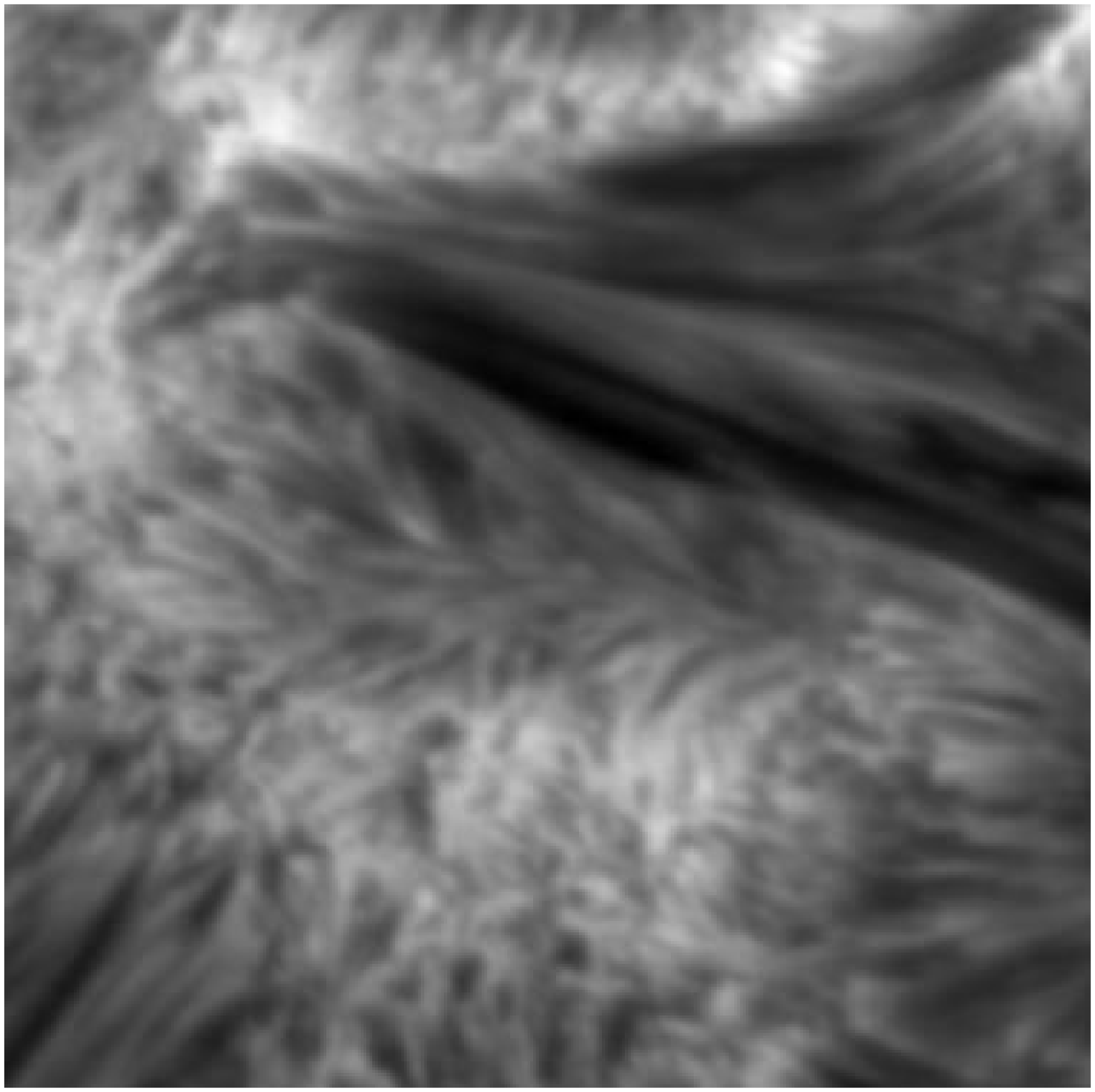}
%\caption{fig2}
\end{minipage}%
}%
\subfigure[ML Deconvolved Image]{
\begin{minipage}[t]{0.2\linewidth}
\centering
\includegraphics[width=1\textwidth]{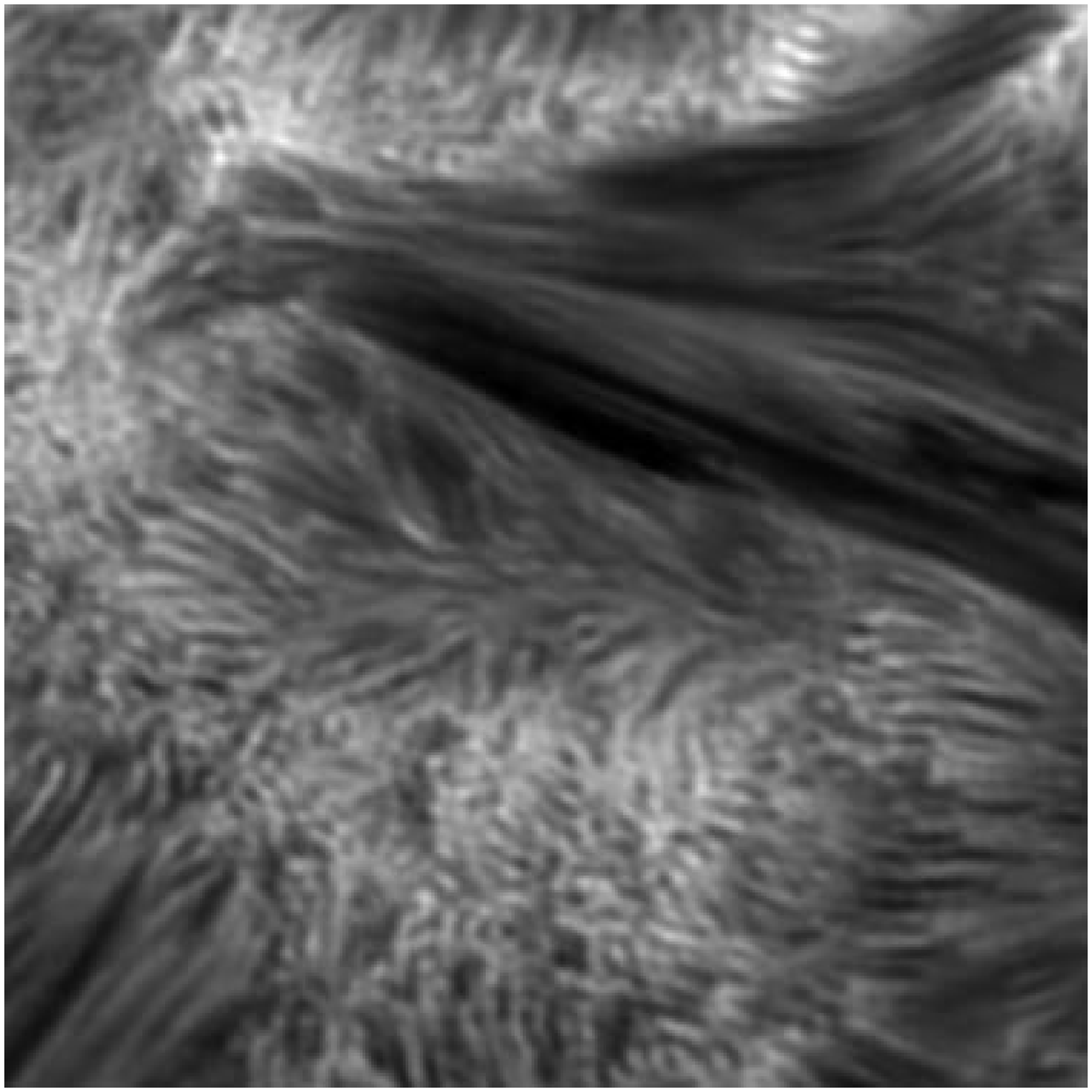}
%\caption{fig2}
\end{minipage}
}%
\subfigure[PSF-NET Restored Image]{
\begin{minipage}[t]{0.2\linewidth}
\centering
\includegraphics[width=1\textwidth]{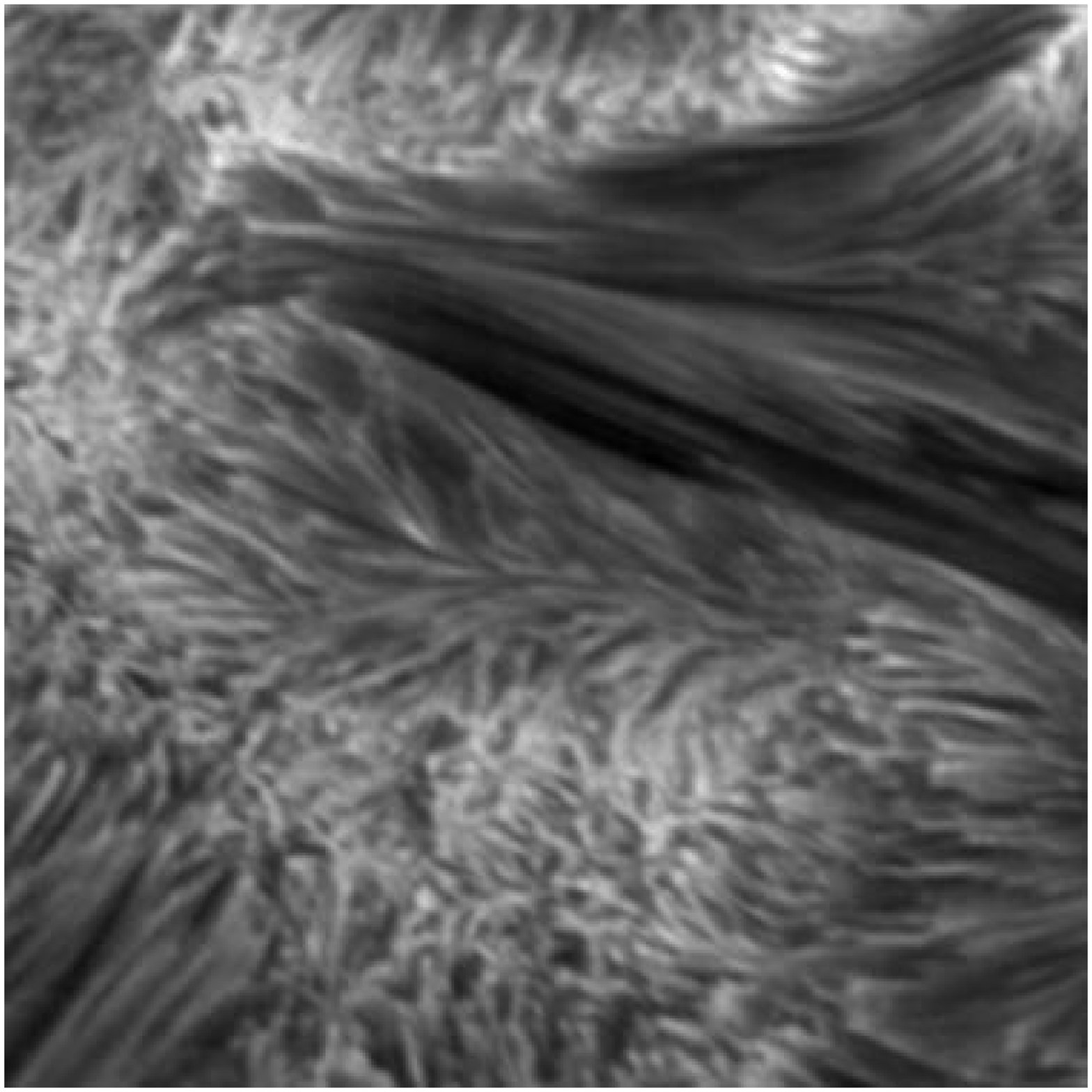}
%\caption{fig2}
\end{minipage}
}%
\centering
\caption{  (a) is the original high resolution H-alpha wavelength image, (b) is the simulated blurred image, (c) is the image restored by the maximum likelihood (ML) blind deconvolution algorithm in the MATLAB and (d) is the image restored by the RESTORE in a trained PSF-NET.}
\label{figure5}
\end{figure*}

\begin{figure*}
\centering
\subfigure[High Resolution Image]{
\begin{minipage}[t]{0.2\linewidth}
\centering
\includegraphics[width=1\textwidth]{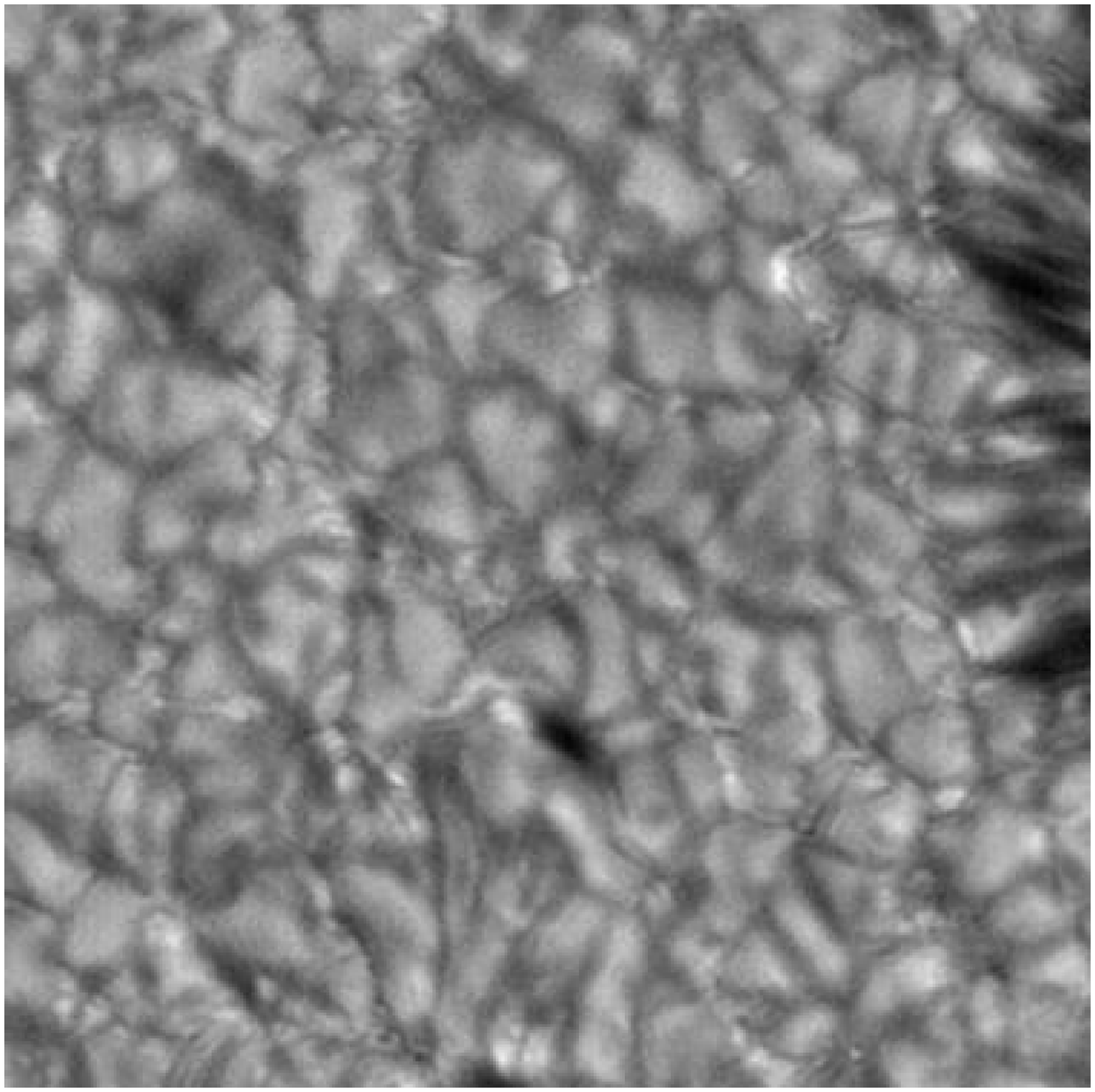}
%\caption{fig1}
\end{minipage}%
}%
\subfigure[Blurred Image]{
\begin{minipage}[t]{0.2\linewidth}
\centering
\includegraphics[width=1\textwidth]{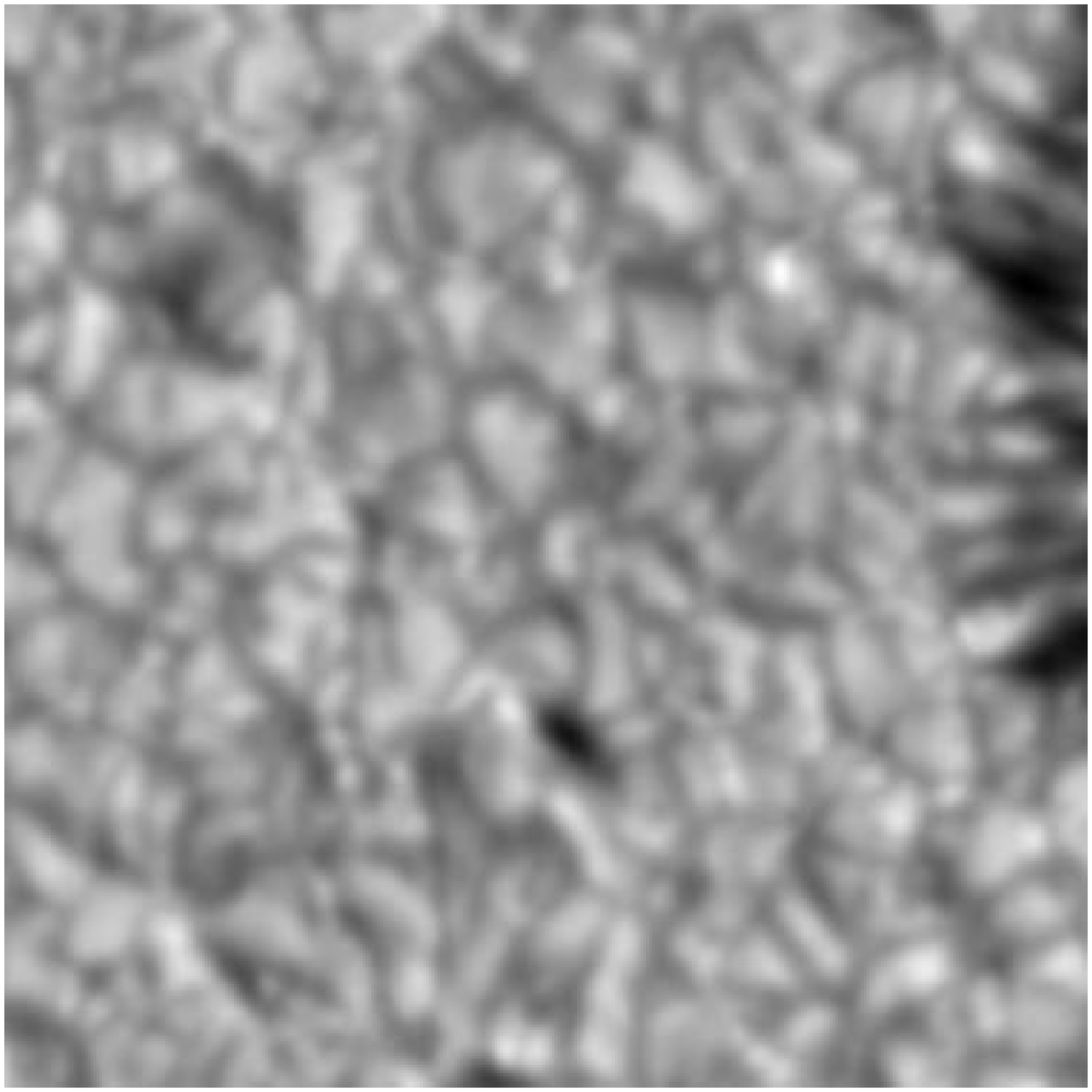}
%\caption{fig2}
\end{minipage}%
}%
\subfigure[ML Deconvolved Image]{
\begin{minipage}[t]{0.2\linewidth}
\centering
\includegraphics[width=1\textwidth]{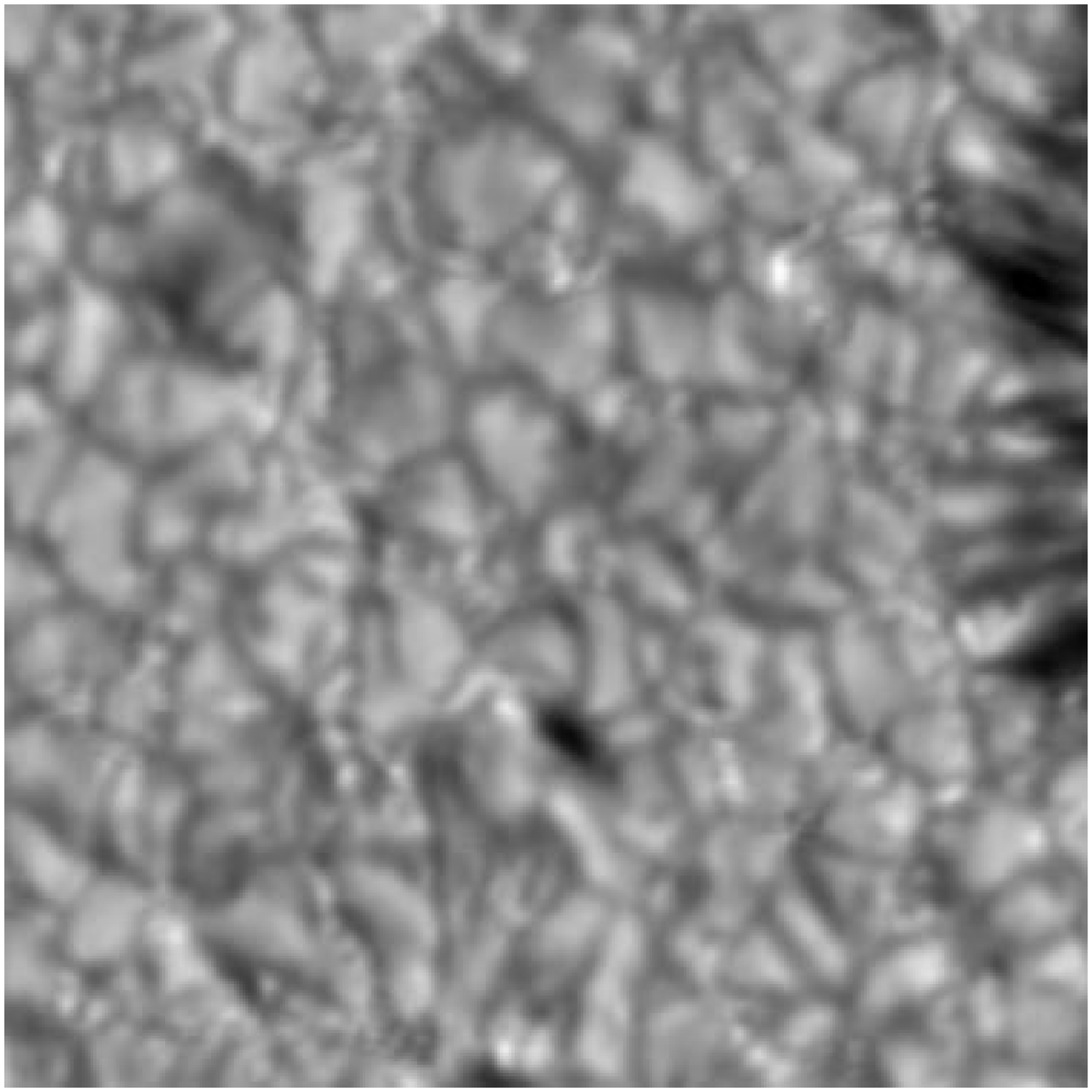}
%\caption{fig2}
\end{minipage}
}%
\subfigure[PSF-NET Restored Image]{
\begin{minipage}[t]{0.2\linewidth}
\centering
\includegraphics[width=1\textwidth]{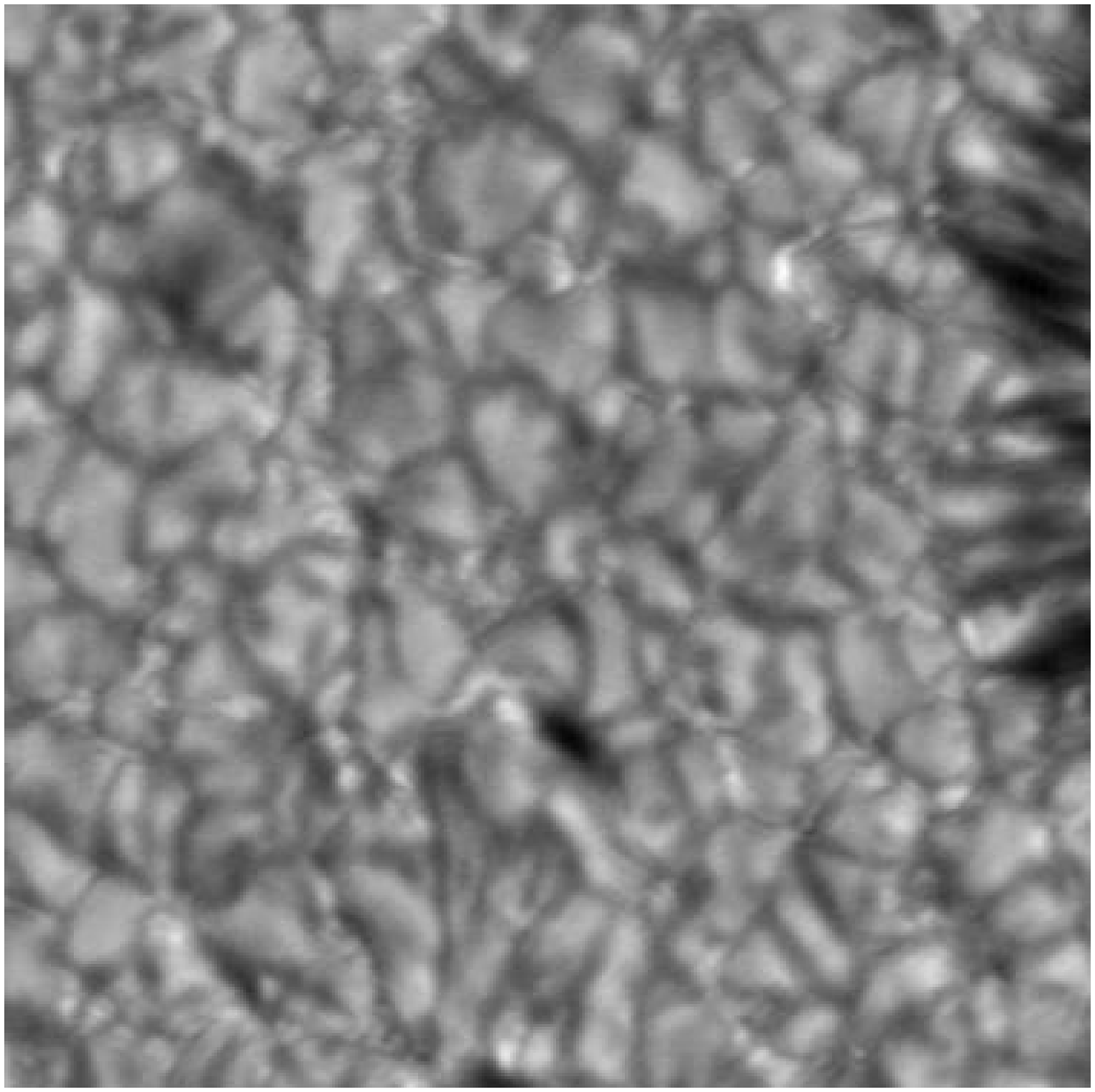}
%\caption{fig2}
\end{minipage}
}%
\centering
\caption{  (a) is the original high resolution H-alpha wavelength image, (b) is the simulated blurred image, (c) is the image restored by the maximum likelihood (ML) blind deconvolution algorithm in the MATLAB and (d) is the image restored by the RESTORE in a trained PSF-NET.}
\label{figure6}
\end{figure*}

\begin{figure*}
\centering
\subfigure[Real Observation Image]{
\begin{minipage}[t]{0.3\linewidth}
\centering
\includegraphics[width=1\textwidth]{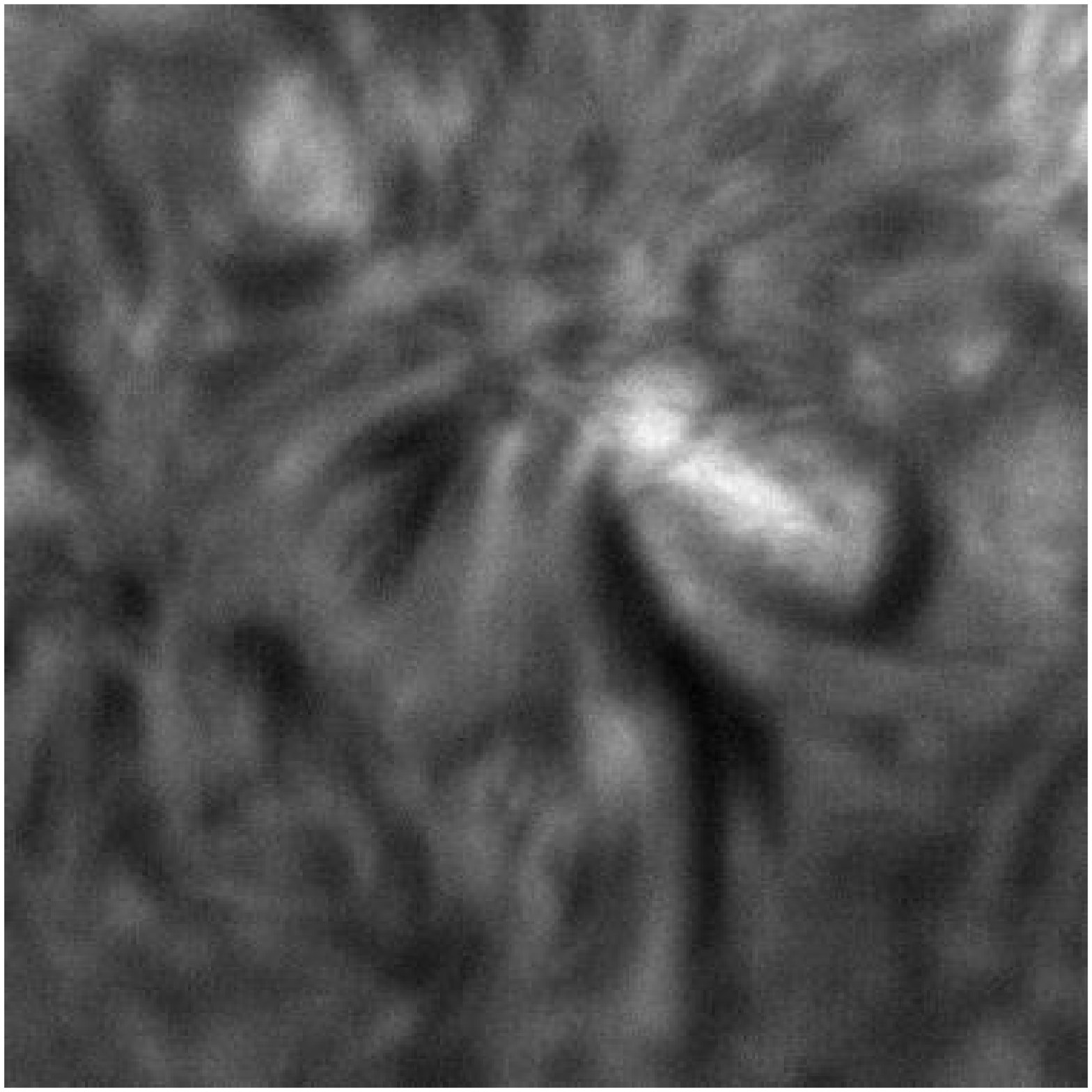}
%\caption{fig1}
\end{minipage}%
}%
\subfigure[ML Deconvolved Image]{
\begin{minipage}[t]{0.3\linewidth}
\centering
\includegraphics[width=1\textwidth]{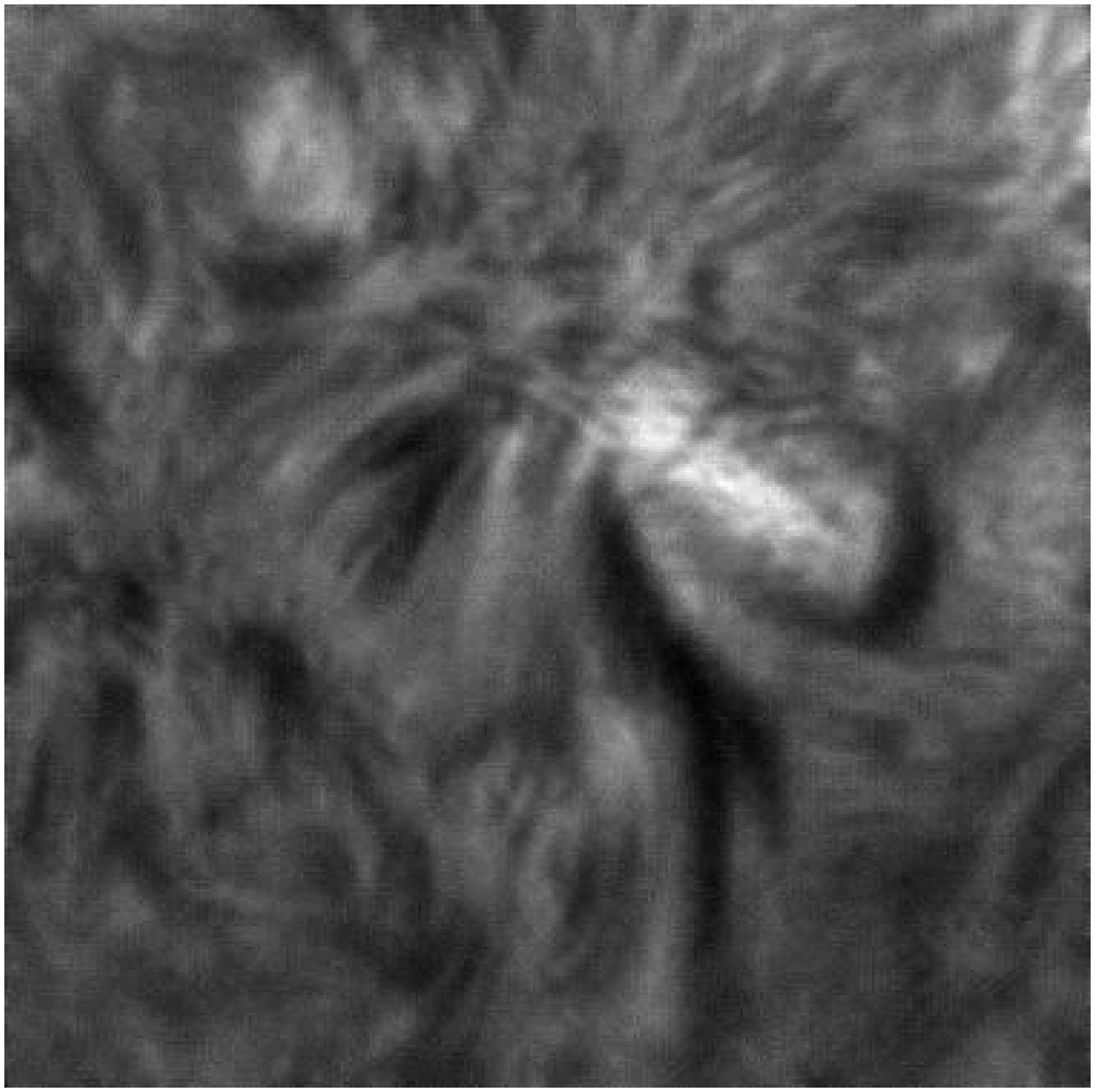}
%\caption{fig2}
\end{minipage}%
}%
\subfigure[PSF-NET Restored Image]{
\begin{minipage}[t]{0.3\linewidth}
\centering
\includegraphics[width=1\textwidth]{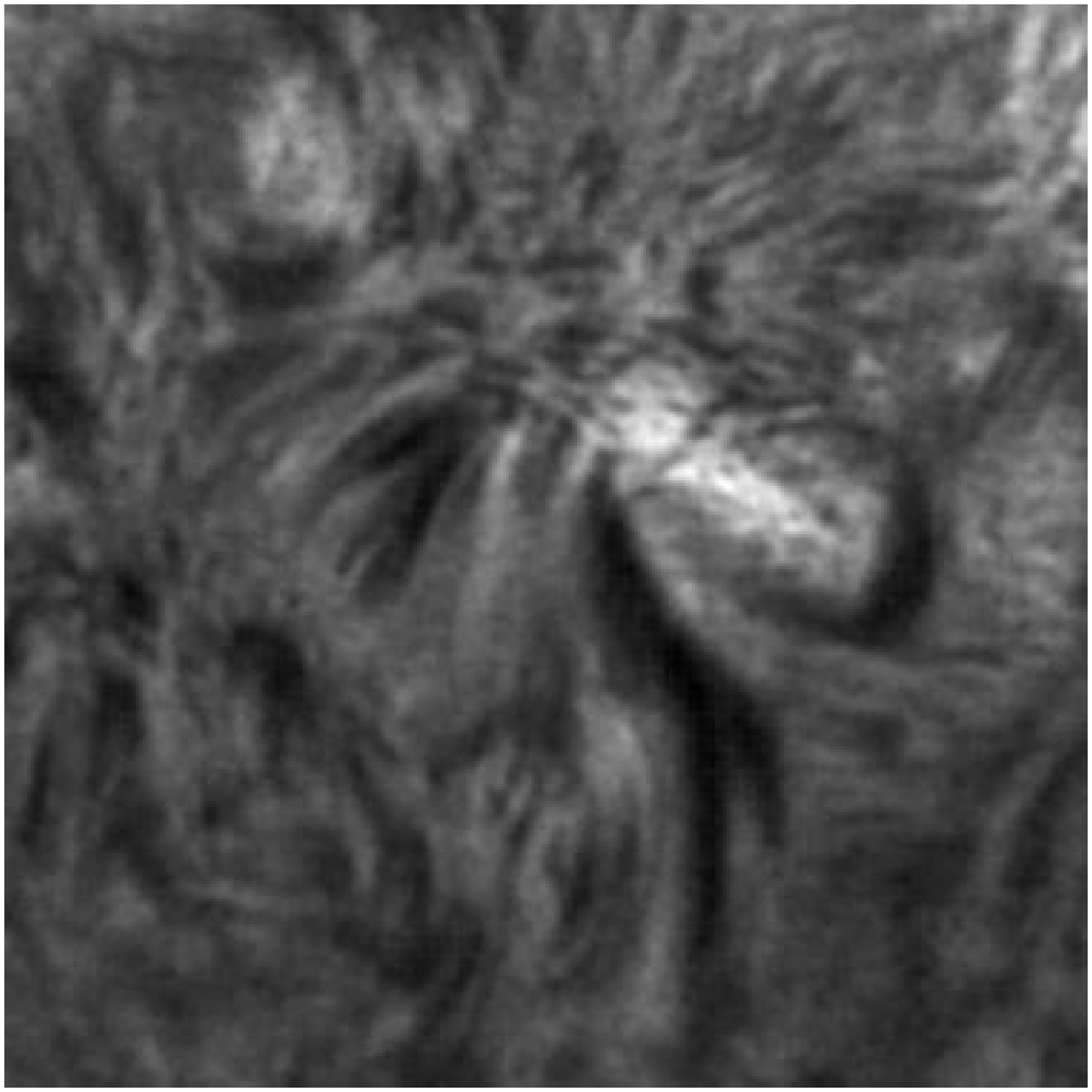}
%\caption{fig2}
\end{minipage}%
}%
\centering
\caption{ (a) is the real observation H-alpha wavelength image, (b) is the image restored by the maximum likelihood (ML) blind deconvolution algorithm in the MATLAB and (c) is the image restored by the RESTORE in a trained PSF-NET.}
\label{figure7}
\end{figure*}

We input images from the test set into the RESTORE neural network in a trained PSF-NET to test its generalization ability. These images are simulated H-alpha wavelength images and TiO band images blurred by PSFs which are generated by Monte Carlo models with the same turbulence profile and different random numbers. We find that restored images are almost the same as the original high resolution images. Particularly, although the PSF-NET is trained with H-alpha images, no filament-like artifacts are introduced into restored TiO band images. Besides, we use the classical maximum likelihood (ML) blind deconvolution algorithm in MATLAB to restore blurred images as comparison. The restored images and the original high resolution images are shown in figure \ref{figure5} and figure \ref{figure6}. We also calculate the peak signal-to-noise ratio (PSNR) of images restored by these two methods. The PSNR is defined by:\\
%--------------------------------------------------------------------
\begin{equation} \label{eq:equation6}
PSNR=\frac{1}{mean [Image_0(x_i,y_i)-Image_1(x_i,y_i)]},
\end{equation}
%-------------------------------
where $Image_0$ and $Image_1$ are two Images and $x_i, y_i$ stand for coordinate in Images. If $Image_0$ is equal to $Image_1$, we directly set $mean [Image_0(x_i,y_i)-Image_1(x_i,y_i)]$ to be 1. We use 100 images blurred by 100 different PSFs as test set  (50 H-alpha wavelength images and 50 TiO band images). Then we use the trained PSF-NET and the ML blind deconvolution algorithm to restore these images. The results are shown in table \ref{table4}. Obviously, the average PSNR of PSF-NET is higher than that of traditional ML method. These tests show that a trained PSF-NET could represent short exposure PSFs generated by the same turbulence profile. To further test the robustness of PSF-NET, we use our method to restore real observation image from the NVST. We estimate the $r_0$ as  from this image according to our experience. Then we generate simulated PSF to train the PSF-NET. The restored image is shown in figure \ref{figure7} and we can find that the PSF-NET has the ability to restore the blurred image.\\
%--------------------------------------------------------------------
	\begin{table}
	\caption{Average PSNR of restoration results}                 
	\centering          
	\begin{tabular}{c c}     % 2 columns 
	\hline\hline       
    % To combine 4 columns into a single one 
	Name  &  Average PSNR\\
	\hline                    
	Blurred Image & 58.83\%\\ 
	ML Deconvolved Image & 60.64\%\\
	PSF-NET Restored Image & 62.31\%\\
	\hline                  
	\end{tabular}
	\label{table4}
	\end{table}
%--------------------------------------------------------------------
%--------------------------------------------------------------------
%--------------------------------------------------------------------
%--------------------------------------------------------------------
%--------------------------------------------------------------------
\subsection{The Interpretation Property of the PSF-NET}
Although most applications of DNNs assume the DNN as a black box, interpretation of the DNN is quite important, if the DNN is related to real physical process or we need to understand its effectiveness \citep{Montavon2017,Huang2019}. Because in this paper, we try to use the PSF-NET to model the manifold space of PSFs induced by the atmospheric turbulence with the same profile and we need to output PSFs for other scientific observations, such as PSF fitting astrometry or photometry, we need to know what the PSF-NET actually learns.\\

\begin{figure}
   \centering
   \includegraphics[width=0.3\hsize]{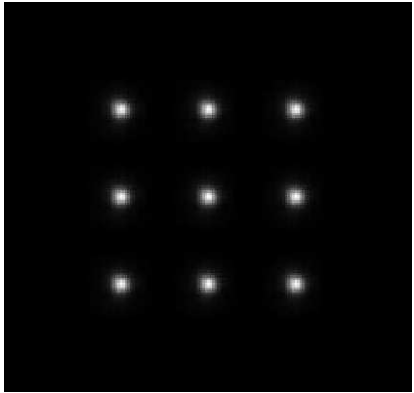}
      \caption{PSF obtained by impulse response at different locations. 
              }
   \label{figure8}
   \end{figure}

Thanks to the Cycle-CNN structure, we can use part of the trained PSF-NET to extract information learned by it. The information learned by the RESTORE neural network are too abstract to understand, while that learned by the PSF neural network should be the PSF. In this part, we plan to interpret information learned by the PSF neural network. The PSF is defined as response of the optical system to a point. If we input an image with pulse signal into the PSF neural network, the output of the PSF neural network should be the PSF. Based on this principle, we input an impulse image which is 1 in some pre-defined positions and 0 in the rest positions into the PSF neural network of a trained PSF-NET. Then we could obtain response of the PSF neural network in different positions as shown in figure \ref{figure8}. In this figure, we can find that the response of the PSF neural network in the PSF-NET is the same and they satisfy spatial invariant property.\\

\begin{figure}
   \centering
   \includegraphics[width=0.5\hsize]{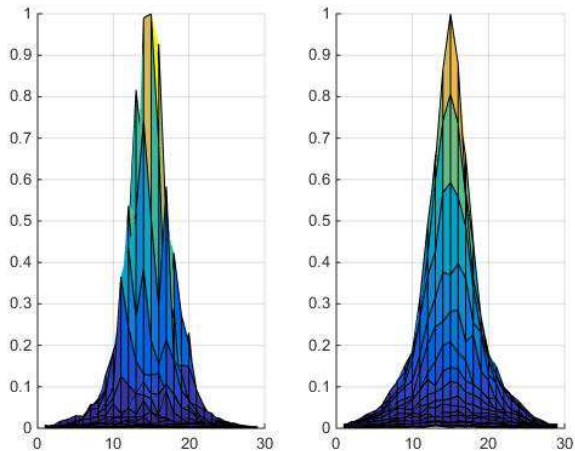}
      \caption{Projection of 3D plot of PSFs. The left figure is a PSF in training set and the right figure is the response of the PSF neural network in the PSF-NET. They have similar scale and shape.}
   \label{figure9}
   \end{figure}

Besides, considering convolution of high resolution images and PSFs are used as training set and these PSFs have the same pixel scale, the contents learned by the PSF neural network should have the same pixel scale as that of PSFs used in the training set. We draw the pulse response of the PSF neural network in the PSF-NET and one of PSFs used in the training set in figure \ref{figure9}. We can find that the size and shape of the pulse response of the PSF neural network in the PSF-NET is close to PSFs in training set. The pulse response of the PSF neural network plays a similar role as `statistical mean PSF', albeit it is obtained by a different way.\\

\begin{figure*}
\centering
\subfigure[High resolution image]{
\begin{minipage}[t]{0.3\linewidth}
\centering
\includegraphics[width=1\textwidth]{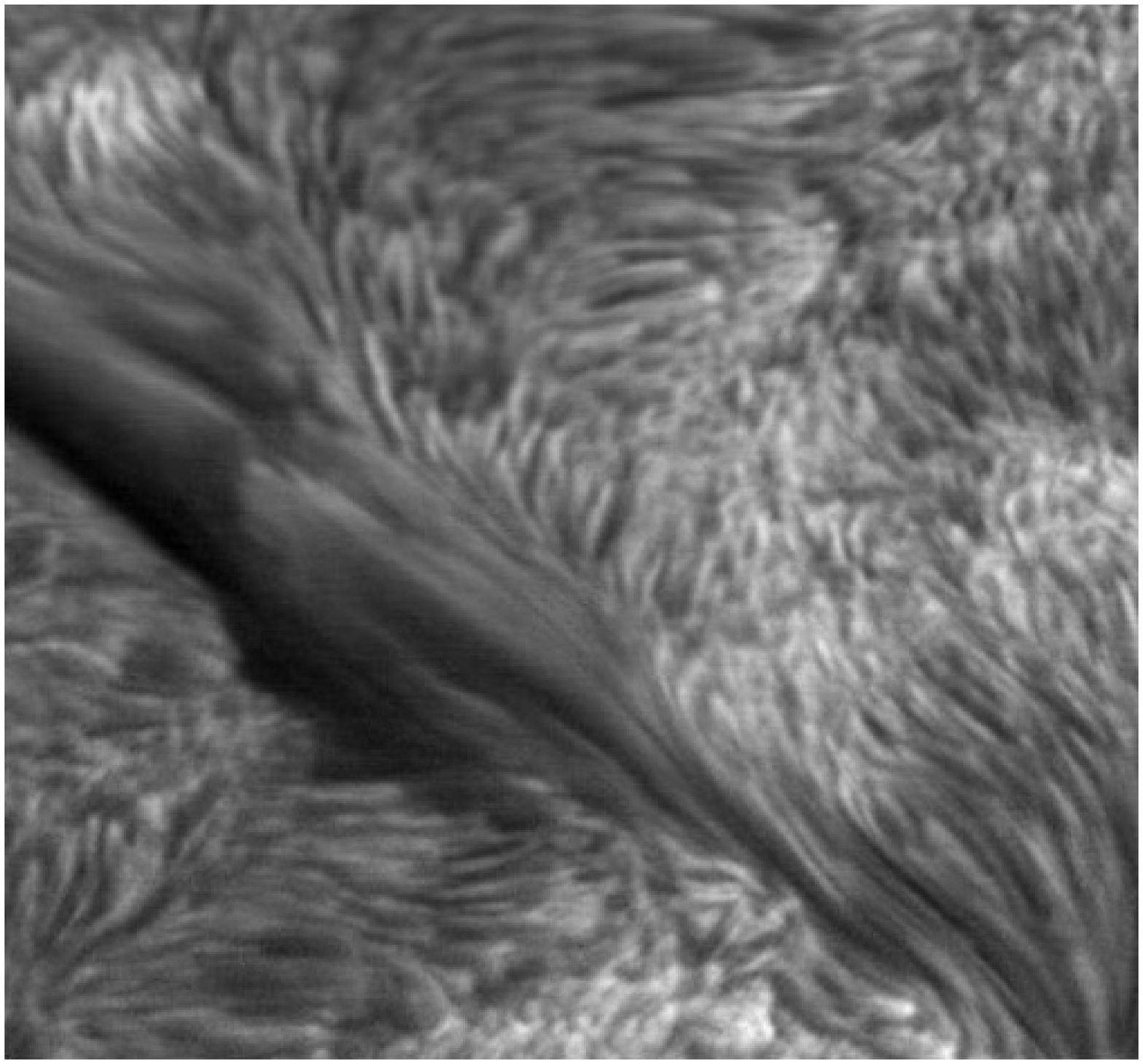}
%\caption{fig1}
\end{minipage}%
}%
\subfigure[Image convolved by the response of the PSF network]{
\begin{minipage}[t]{0.3\linewidth}
\centering
\includegraphics[width=1\textwidth]{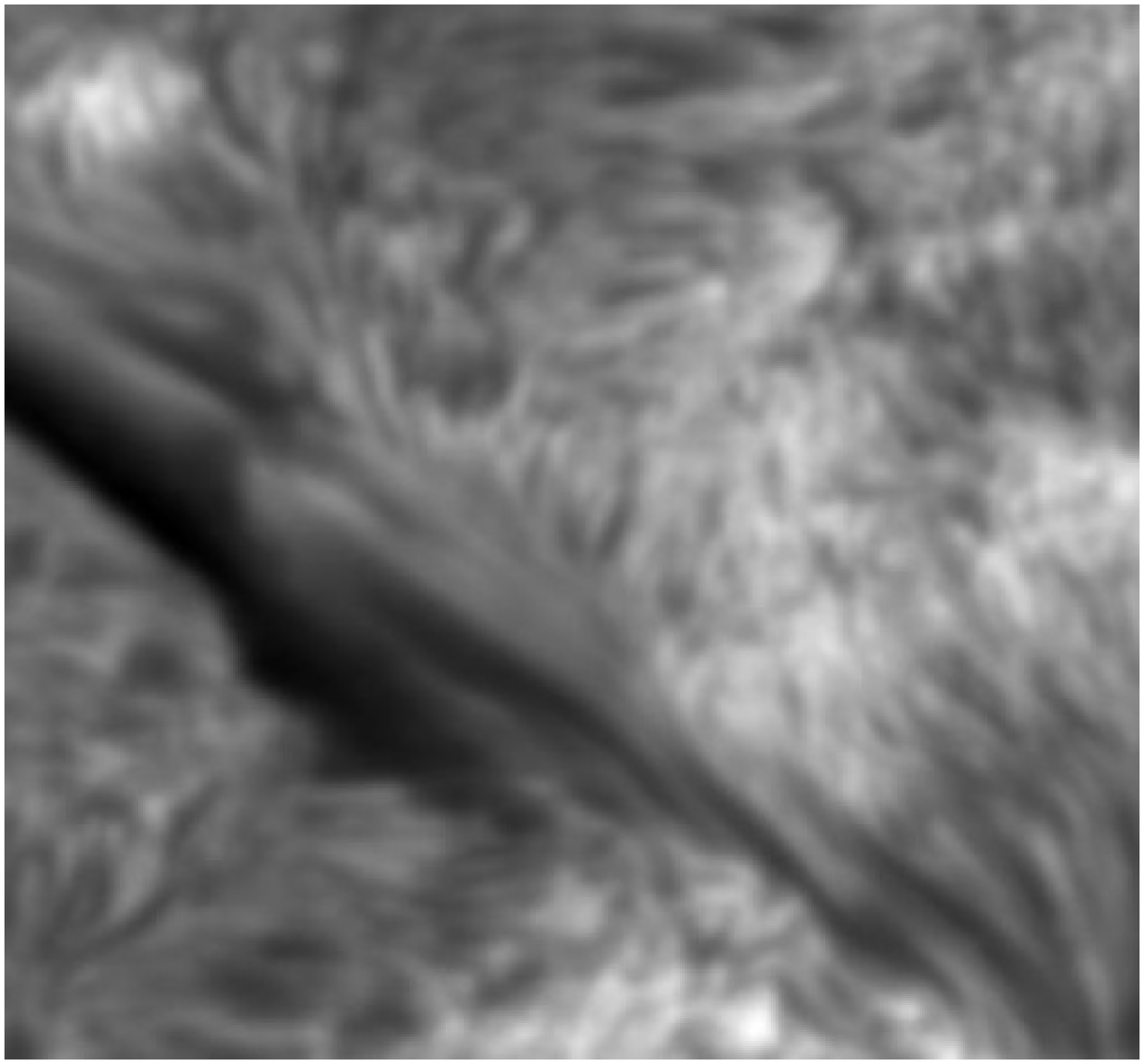}
%\caption{fig2}
\end{minipage}%
}%
\subfigure[Image restored by the RESTORE network]{
\begin{minipage}[t]{0.3\linewidth}
\centering
\includegraphics[width=1\textwidth]{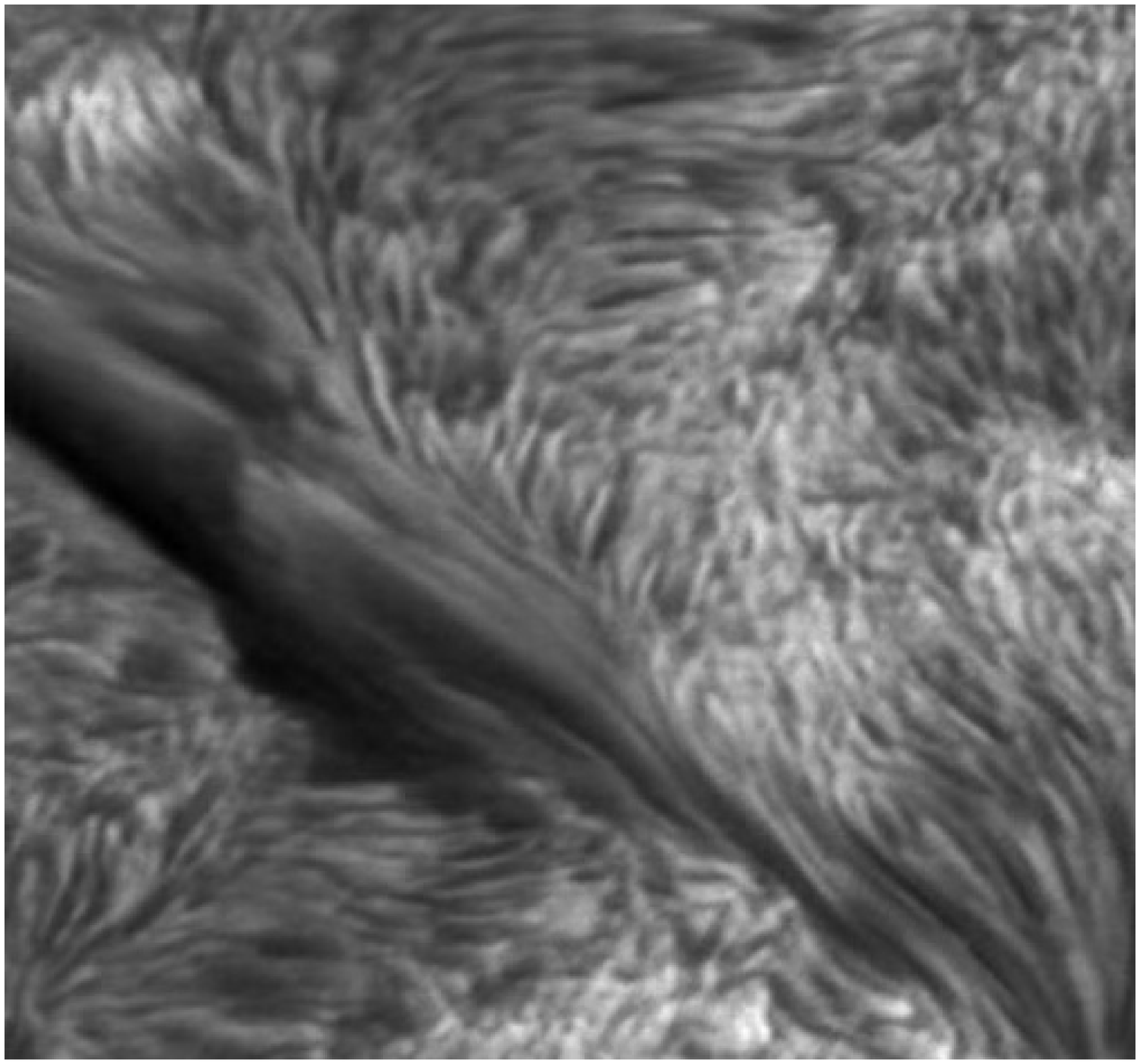}
%\caption{fig2}
\end{minipage}
}%
\centering
\caption{ (a) is the high resolution H-alpha wavelength image, (b) is convolution result of high resolution image and the PSF, and (c) is the image restored by the RESTORE neural network.}
\label{figure10}
\end{figure*}

To further study contents learned by the PSF-NET, we focus on the RESTORE neural network in the PSF-NET, since the RESTORE neural network plays a similar role as deconvolution kernel and the PSF neural network plays a similar role as the PSF. Images blurred by the PSF from the PSF neural network should be successfully restored by the deconvolution kernel from the RESTORE neural network. Based on this idea, we extract the response of the PSF neural network and convolve it with a high resolution image to obtain a blurred image. Then we use the RESTORE neural network to restore the blurred image. The results are shown in figure \ref{figure10}. The results show that the RESTORE neural network and the PSF neural network play the similar role as PSF and deconvolution kernel.\\

In this part, we design three experiments to show the interpretation property of the PSF-NET. Results of these experiments show that the PSF-NET has provided a way for us to analyze properties of atmospheric turbulence induced aberrations. In the next section, we will use PSF-NET to analyze PSF variations caused by atmospheric turbulence variations.\\
%--------------------------------------------------------------------
%--------------------------------------------------------------------
%--------------------------------------------------------------------
%--------------------------------------------------------------------
%--------------------------------------------------------------------
%%%%%%%%%%%%%%%%%%%%%%%%%%%%%%%%%%%%%
\section{Analyzing PSF variations with the PSF-NET}\label{sec:analys}
Ground based optical telescopes have PSFs with strong variations both in spatial domain and temporal domain, which is mainly caused by the atmospheric turbulence. In our previous paper, we find that the characteristic time of the atmospheric turbulence profile ($C_n^2$  versus height) variation \citep{Jia2018} has the same magnitude as that of PSFs obtained by the SDSS \citep{Xin2018}. This phenomenon indicates us that different atmospheric turbulence profiles may introduce PSFs that belong to different manifold spaces. But how atmospheric turbulence profiles could introduce different PSFs? Because the pulse response of the PSF neural network in the PSF-NET could model the manifold space of PSFs, we can use it to analyze PSF variations caused by the atmospheric turbulence profile variations.\\

\subsection{Variations of integrated $C_n^2$ and its impacts to PSF variations}
The most distinct impact of atmospheric turbulence profile variations that would introduce PSF variations is variations of the integrated $C_n^2$. Different integrated $C_n^2$ would lead to different coherent length $r_0$. This impact is well known and we firstly analyze this variation with the PSF-NET in this part. We generate 500 PSFs through 20 Monte Carlo simulations with the same turbulence profile as shown in table \ref{table2} and different $D/r_0$ between 10 and 14. These PSFs are used to train the PSF-NET and we extract the pulse response of the PSF neural network in the PSF-NET. We calculate the PSNR of these pulse responses in pairs. Comparison results are shown in table \ref{table5}. We can find that manifold space of PSFs is different, when the atmospheric turbulence has different integrated $C_n^2$. As the difference of integrated $C_n^2$ increases, the difference of the manifold space of PSFs also increases.\\
	\begin{table}
	\caption{PSNR between pulse responses of the PSF neural network in the PSF-NET with different $D/r_0$}                 
	\centering          
	\begin{tabular}{c c c c c }     % 5 columns 
	\hline\hline                         
	$D/r_0$ & 11&12 & 13&14 \\ 
	\hline
	10 &70.52\%&66.48\% & 63.69\%&61.04\% \\                    
	11 & 100\%&66.78\% & 64.44\%&61.32\%  \\                    
	12 & &100\% &65.68\%& 63.34\%  \\  
	13 & & &100\%& 66.84\%  \\
	\hline               
	\end{tabular}
	\label{table5}
	\end{table}
%--------------------------------------------------------------------
\subsection{Variations of $C_n^2$ and its impacts to PSF variations}
We further analyse variations of PSFs brought by variations of the $C_n^2$. We analyse two scenes of $C_n^2$ variations: increasing number of equivalent layers in the atmospheric turbulence profile and using the  same number of equivalent layers with different integrated $C_n^2$.\\

For the first scene, we set D/r0 to 10 and use three turbulence profiles in table \ref{table6} to generate 500 PSFs through 20 Monte Carlo simulations for each profile. With these PSFs, we train the PSF-NET and extract pulse response of the PSF neural network. The results are shown in table \ref{table7}. We can find that as the number of equivalent increases, the difference between turbulence profile also increases, which lead to increased difference between pulse responses of PSFs.\\

In the second scene, we use different atmospheric turbulence profiles as shown in table \ref{table6} to generate PSFs. All parameters are the same as shown in table \ref{table1} and D/r0 is 10. 500 PSFs are generated by 20 Monte Carlo simulations for each profiles and we use these PSFs to train the PSF-NET. We compare PSNR of pulse responses of the PSF neural network in the PSF-NET and the results are shown in table \ref{table8}. We can find that the PSNR is biggest for the PSNR between the turbulence profile 3-1 and the turbulence profile 3-2, because they have three layers and their $C_n^2$ are the most similar. As the difference of atmospheric turbulence profile increases, the PSNR reduces which indicates difference between PSFs increases.\\
%--------------------------------------------------------------------
	\begin{table}
	\caption{The atmospheric turbulence profiles used in this section}                 
	\centering          
	\begin{tabular}{c c}     % 7 columns 
	\hline\hline       
    % To combine 4 columns into a single one 
	Turbulence Profiles  &  Parameters\\
	\hline                    
	1 & [ 1.00 , 0.00 , 0.00 , 0.00 ]\\ 
	2 & [ 0.50 , 0.50 , 0.00 , 0.00 ]\\
	3-1 & [ 0.33 , 0.33 , 0.34 , 0.00 ] \\   
	3-2 & [ 0.30 , 0.30 , 0.40 , 0.00 ] \\  
	4 & [ 0.40 , 0.20 , 0.20 , 0.20 ] \\
	\hline 
	\hline                    
	Height in meter & [ 500 , 1000 , 2000 , 4000 ] \\
	\hline                    
	\end{tabular}
	\label{table6}
	\end{table}
%-------------------------------------------------------------------------
	\begin{table}
	\caption{PSNR between pulse responses of the PSF neural network in the PSF-NET with different equivalent layers}                 
	\centering          
	\begin{tabular}{c c c c}     % 7 columns 
	\hline\hline                         
	Turbulence Profiles &2&3-1 &4 \\
	\hline   
	1 &68.48\%&66.45\% & 61.86\% \\                    
	2 &100\%&68.88\% & 62.72\%  \\                    
	3-1 & & 100\% & 65.05\%  \\
	\hline                
	\end{tabular}
	\label{table7}
	\end{table}
%--------------------------------------------------------------------
	\begin{table}
	\caption{PSNR between pulse responses of the PSF neural network in the PSF-NET with different turbulence profiles}                 
	\centering          
	\begin{tabular}{c c c c c}     % 7 columns 
	\hline\hline                         
	Turbulence Profiles &1&2&3-1 &4 \\
	\hline   
	3-2 &66.54\%&67.90\% &70.14\%& 65.66\% \\
	\hline                
	\end{tabular}
	\label{table8}
	\end{table}
%%%%%%%%%%%%%%%%%%%%%%%%%%%%%%%%%%%%%
\section{Conclusions and Future Work}\label{sec:con}
This paper proposes a DNN based PSF model PSF-NET for ground based optical telescopes. The PSF-NET is an interpretable DNN model, which is trained with Monte Carlo simulated PSFs. After training, it can learn manifold space of PSFs that are generated by the same atmospheric turbulence profile. Images blurred by PSFs generated by the same atmospheric turbulence profile can be effectively restored by the RESTORE neural network in the PSF-NET. Furthermore, the pulse response of the PSF neural network in the PSF-NET is the statistical mean PSF which can reflect PSF variations brought by atmospheric turbulence profile variations. The PSF-NET is a useful tool to further analyze atmospheric turbulence induced aberrations, which would be helpful for new image restoration methods design or new adaptive optics system development.\\

However, it should be noted that successful applications of the PSF-NET depend on high accuracy measurements of atmospheric turbulence profile and high fidelity Monte Carlo simulation. Errors propagate from turbulence measurement or deviation from the theoretical model \citep{Martinez2010} would still limit its performance. To reduce impacts brought by these uncertainties and achieve PSF model with high accuracy, we need to introduce additional regularization conditions, such as wavefront measurements or real observation images. We will carry on our research in this area in our future papers. Besides, we will use real measurements of turbulence profile and observed PSF to verify the PSF-NET in our future work.\\

\section*{Acknowledgments}
The authors would like to thank the anonymous referee for comments and suggestions that greatly improved the quality of this manuscript. The authors would like to thank Dr. Yongyuan Xiang from Yunnan Astronomical Observatory for providing solar observation data from NVST to this paper. P.J. would like to thank Professor Hui Liu, Professor Kaifan Ji and Professor Zhong Liu form Yunnan Astronomical Observatory, Dr. James Osborn, Dr. Alastair Basden, Dr. Tim Morris and Dr. Matthew Townson from Durham University, Professor François Rigaut from Australian National University who provide very helpful suggestions for this paper. This work is supported by National Natural Science Foundation of China (NSFC)(11503018), the Joint Research Fund in Astronomy (U1631133) under cooperative agreement between the NSFC and Chinese Academy of Sciences (CAS), Shanxi Province Science Foundation for Youths (201901D211081), Research and Development Program of Shanxi (201903D121161), Research Project Supported by Shanxi Scholarship Council of China, the Scientific and Technological Innovation Programs of Higher Education Institutions in Shanxi (2019L0225).\\

The data used in this paper were obtained with the New Vacuum Solar Telescope in Fuxian Solar Observatory of Yunnan Astronomical Observatory, CAS. The code used in this paper is written in Python programming language (Python Software Foundation) with the package pytorch and can be downloaded from \href{JPwebsite}{http://aojp.lamost.org}\\

\bibliographystyle{aasjournal}
\bibliography{psfnet}

%% thebibliography produces citations in the text using \bibitem-\cite
%% cross-referencing. Each reference is preceded by a
%% \bibitem command that defines in curly braces the KEY that corresponds
%% to the KEY in the \cite commands (see the first section above).
%% Make sure that you provide a unique KEY for every \bibitem or else the
%% paper will not LaTeX. The square brackets should contain
%% the citation text that LaTeX will insert in
%% place of the \cite commands.

%% We have used macros to produce journal name abbreviations.
%% \aastex provides a number of these for the more frequently-cited journals.
%% See the Author Guide for a list of them.

%% Note that the style of the \bibitem labels (in []) is slightly
%% different from previous examples.  The natbib system solves a host
%% of citation expression problems, but it is necessary to clearly
%% delimit the year from the author name used in the citation.
%% See the natbib documentation for more details and options.

%% This command is needed to show the entire author+affilation list when
%% the collaboration and author truncation commands are used.  It has to
%% go at the end of the manuscript.
%\allauthors

%% Include this line if you are using the \added, \replaced, \deleted
%% commands to see a summary list of all changes at the end of the article.
%\listofchanges

\end{document}